\documentclass[
reprint,
superscriptaddress,
 aps,
pra]{revtex4-2} 
\usepackage{lipsum} 
\usepackage{comment,color,xcolor,amsmath,amssymb}
\usepackage{graphicx}
\usepackage{dcolumn}
\usepackage{bm}

\begin{document}

\preprint{APS/123-QED}

\title{On a time-resolved interpretation of the Husimi function} 

\author{Ralph Sabbagh}
\affiliation{Department of Mechanical and Aerospace Engineering, University of California, Irvine, California 92697, USA}
\author{Olga Movilla Miangolarra} 
\affiliation{Department of Physics and Instituto Universitario de Estudios Avanzados (IUdEA),\\ Universidad de La Laguna, La Laguna 38203, Spain}
\author{Tryphon T. Georgiou}
\affiliation{Department of Mechanical and Aerospace Engineering, University of California, Irvine, California 92697, USA}


\begin{abstract}
In this Letter, we interpret the Husimi function as the conditional probability density of continuously measuring a stream of constant position and momentum outcomes, indefinitely.
This gives rise to an alternative definition that naturally extends to an arbitrary collection of self-adjoint operators without reference to coherent states.
This definition recovers the Husimi distribution for a spin-half particle when monitoring the three Pauli matrices, as well as Born's rule for quantum measurement when monitoring commuting quantum observables.
%
%
Ultimately, the proposed paradigm generates positive representations of quantum states as conditional densities, on both finite and infinite time classical experiments, as expectations of a fundamental operator, the Gaussian semigroup.
%
\end{abstract}

\maketitle


The theory of quantum measurement posits a natural duality between information and time, in which one may conceive of the Von Neumann collapse of a system to an observable's eigenstate as the result of performing either
\begin{itemize}
\setlength{\itemindent}{-1em}
    \item[i)] a projective measurement that instantly extracts all the information about the observable, or
    \item[ii)] a continuous measurement that gradually siphons this information out, over an infinite time horizon \footnote{This refers to an ideal (diffusive) continuous measurement of some non-degenerate observable in the absence of a Hamiltonian.}.
\end{itemize}

In either scenario, the eventual collapse can be seen as dictated by the \textit{average value} of the collected measurements--- whether it be a single outcome (as in i) or infinitely many (as in ii). In both cases, this value follows \textit{the same probability distribution}. Thus, from a black-box perspective, if all that is recorded is this average value and the collapsed state, the two scenarios are indistinguishable.

The above equivalence breaks down when measuring \textit{non-commuting} observables. 
The first scenario is no longer admissible since it is impossible, for instance, to know a quantum system's position and momentum (corresponding to the observables $\hat X$ and $\hat P$) simultaneously. In contrast, the second scenario can still be realized as it does not violate the uncertainty principle at the outset; the (forbidden) simultaneous measurement is now resolved along the infinite time-axis. 

A manifestation of this dichotomy shows up when attempting to write a joint probability density on $\mathbb R^n$ for a collection of $n$ possibly non-commuting observables under scenario i).
For instance, in the case of two non-commuting observables $\hat X$ and $\hat P$, the {\em Wigner distribution} $W_\rho$ is in general sign indefinite 
 \cite{wigner1932quantum},~\footnote{The {\em Wigner distribution} $W_\rho$ is the Fourier transform of the characteristic function $\chi(\xi)=\text{tr}(\rho e^{i\xi_1 \hat X + i\xi_2 \hat P})$.}.
The purpose of this Letter is to show that adopting scenario ii) instead allows positive representations of quantum states for an arbitrary number of non-commuting observables; for $\hat X$ and $\hat P$, the representation relates to the well-known {\em Husimi function} which is a Gaussian smearing (convolution) of $W_\rho$ \cite{husimi1940some}.

In what follows, we shall first prove that for an initial state $\rho$, the joint probability of obtaining average values of $x$ and $p$ from continuous measurement over a time window $[0,\tau]$, conditioned on observing a stream of \textit{constant} outcomes, is given by
\begin{subequations}
\begin{align}
   \hspace*{-5pt} H_{\tau,\rho}(x,p)&=\cfrac{\text{tr}\left(\rho e^{-\tau((\hat{X}-x)^2+(\hat{P}-p)^2)}\right)}{\int_{\mathbb{R}^2}\text{tr}\left(\rho e^{-\tau((\hat{X}-x')^2+(\hat{P}-p')^2)}\right)\text{d}x'\text{d}p'}.\label{eq:1}
\end{align}
This expression is manifestly positive for any $\rho$ and integrates to one. In fact, it is exactly the convolution
\begin{align}\label{eq:conv}
    H_{\tau,\rho}(x,p)&=\left(W_\rho \ast g_{\sigma_\tau}\right)(x,p),
\end{align}
\end{subequations}
where $g_{\sigma_\tau}$ is a bivariate Gaussian with variance
\begin{align*}
    \sigma^2_\tau :=\frac{\hbar}{2\tanh(\hbar \tau)}. 
\end{align*}
This, as time goes to infinity, is none other than the
Husimi function \footnote{An alternative expression for the Husimi function is given by $H_\rho(\alpha)=\langle \alpha |\rho|\alpha\rangle/\pi$, where $|\alpha\rangle$ is a coherent state corresponding to $\alpha=(x+ip)/\sqrt{2\hbar}$.}
\begin{align*}
     H_{\infty,\rho} :=W_{\rho}\ast g_{\hspace*{-2pt}\,_{\tiny\sqrt{\hbar/2}}},
\end{align*}
and therefore provides the probability law for the time-resolved simultaneous measurement of the position and momentum observables described in scenario ii), conditioned on monitoring a stream of constant position and momentum outcomes indefinitely.

The significance of this interpretation is three-fold.
First, it formulates the Husimi function in terms of continuous measurements {\em without reference to coherent states}; notably, it trivializes the reasoning that explains why the Gaussian smearing of the Wigner distribution in \eqref{eq:conv} is manifestly positive for any $\rho$. Second, expression \eqref{eq:1} and its interpretation in terms of continuous measurements readily extend to an arbitrary collection of non-commuting observables $\hat A_1,\ldots,\hat A_n$, namely,
\begin{subequations}
\begin{align}\label{eq:Z}
    H_{\tau,\rho}(a) = \cfrac{\text{tr}\left(\rho e^{-\tau{\sum_{k=1}^n(\hat{A}_k-a_k)^2}}\right)}{\int_{\mathbb{R}^n}\text{tr}\left(\rho e^{-\tau{\sum_{k=1}^n(\hat{A}_k-a^{'}_k)^2}}\right)\text{d}a^{'}},
\end{align}
giving rise to a positive representation of $\rho$ in $\mathbb R^n$, and thereby, suggests a corresponding generalization of the Husimi function.
Third, the correspondence between \eqref{eq:1} and \eqref{eq:conv},
extends to one between \eqref{eq:Z} and
\begin{align}\label{eq:general}
    H_{\tau,\rho}(a)&=\int_{\mathbb R^n} W_\rho(a') \frac{g_{\tau,a}(a')}{Z_{\tau,\rho}}da',
\end{align}
\end{subequations}
which provides
the {\em natural smearing operation} of the Wigner distribution that eliminates negative values in this generality; here $W_\rho$ is the generalized Wigner distribution \cite{schwonnek2020wigner,sabbagh2025particle} with respect to $\hat A_1,\ldots,\hat A_n$ and
$g_{\tau,a}$ is the Weyl symbol \cite{anderson1969weyl,anderson1970weyl} of the {\em Gaussian semigroup} $e^{-\tau{\sum_{k=1}^n(\hat{A}_k-a_k)^2}}$ with ${Z_{\tau,\rho}}$ the denominator in \eqref{eq:Z}. 


Interestingly, while the smearing kernel $g/Z$ in \eqref{eq:general} may not be a shifted Gaussian in general, the form of the operator when viewed in Hilbert space, {\em always} is.

We now proceed to derive \eqref{eq:1}, which is the conditional probability density of continuously measuring a stream of constant position and momentum outcomes.
To this end, we recall that a continuous measurement is the limit of
a succession of weak measurements  \cite{jacobs2006straightforward,jordan2024quantum}.
Specifically, weakly measuring position $x$  (resp.\ momentum $p$) channels the initial state $\rho$ 
through
\begin{align*}
    \rho \mapsto \cfrac{\chi^{\tau}_{x}\rho \chi^{\tau}_{x}}{\text{tr}(\chi^{\tau}_{x}\rho \chi^{\tau}_{x})}~~~~~ (\mbox{resp.\ }\rho \mapsto \cfrac{\Xi^{\tau}_{p}\rho \Xi^{\tau}_{p}}{\text{tr}(\Xi^{\tau}_{p}\rho \Xi^{\tau}_{p})}),
\end{align*}
where
    \begin{align*}
     \chi_x^\tau=
     \sqrt[4]{\frac{\tau}{\pi}}
     e^{-\frac{\tau}{2}(\hat{X}-x)^2},
     ~~\Xi_p^\tau=
     \sqrt[4]{\frac{\tau}{\pi}}
     e^{-\frac{\tau}{2}(\hat{P}-p)^2},
\end{align*}
and $\tau$ is the duration of the measurement.
The normalizing factors
\begin{align*}
    \text{tr}(\chi_x^\tau\rho\chi_x^\tau)~\text{  and 
 }~\text{tr}(\Xi_p^\tau\rho\Xi_p^\tau),
\end{align*}
represent the probability of measuring $x$ and $p$, respectively. Alternating between $n$ such position and momentum measurements, with duration $\tau/2n$ each, gives
\begin{align}
    \text{tr}\left(\Xi^{\frac{\tau}{2n}}_{p_n}\chi_{x_n}^{\frac{\tau}{2n}}\ldots\Xi^{\frac{\tau}{2n}}_{p_1}\chi_{x_1}^{\frac{\tau}{2n}}\rho\chi_{x_1}^{\frac{\tau}{2n}}\Xi^{\frac{\tau}{2n}}_{p_1}\ldots\chi_{x_n}^{\frac{\tau}{2n}}\Xi^{\frac{\tau}{2n}}_{p_n}\right)\label{eq:joint}
\end{align}
as the probability of a sequence $(x_1,p_1,\ldots,x_n,p_n)$ of measurement outcomes. This is a classical probability.

Next we compute the probability of obtaining average values $x$ and $p$, \emph{conditioned on the stream of outcomes remaining constant}, i.e., $x_i=x_j$, $p_i=p_j$ for $1\leq i,j\leq n$. From \eqref{eq:joint}, a direct application of Bayes' rule gives
\begin{align*}
    \cfrac{ \text{tr}\left((\Xi^{\frac{\tau}{2n}}_{p}\chi_{x}^{\frac{\tau}{2n}})^n\rho(\chi_{x}^{\frac{\tau}{2n}}\Xi^{\frac{\tau}{2n}}_{p})^n\right)}{\int_{\mathbb{R}^2}\text{tr}\left((\Xi^{\frac{\tau}{2n}}_{p'}\chi_{x'}^{\frac{\tau}{2n}})^n\rho(\chi_{x'}^{\frac{\tau}{2n}}\Xi^{\frac{\tau}{2n}}_{p'})^n\right)\text{d}x'\text{d}p'}.
\end{align*}
Substituting the expressions for $\chi_x$ and $\Xi_p$ and simplifying yields 
\begin{align}
    \cfrac{ \text{tr}\left(M_n(x,p)^\dagger\rho M_n(x,p)\right)}{\int_{\mathbb{R}^2}\text{tr}\left(M_n(x',p')^\dagger\rho M_n(x',p')\right)\text{d}x'\text{d}p'},\label{eq:cond}
\end{align}
where  
$
   M_n(x,p)= (e^{-\frac{\tau}{2n}(\hat{X}-x)^2}e^{-\frac{\tau}{2n}(\hat{P}-p)^2})^n$ and $M^\dagger$ denotes the adjoint of $M$.
   Using the Lie-Trotter formula \cite{CHERNOFF1968238}, we obtain that
\begin{align*}
    \lim_{n\rightarrow\infty}M_n(x,p) = e^{-\frac{\tau}{2}((\hat{X}-x)^2+(\hat{P}-p)^2)}.
\end{align*}
Substituting into $\eqref{eq:cond}$ \footnote{The limit is taken in the strong operator topology and is passed through the integral in the denominator provided \begin{align*}
    \inf_{f\in L_1^+(\mathbb R^2)}
    \sup_{n\in\mathbb N} \iint_{{g_n\geq f}}
     g_n(x,p)\text{d}x\text{d}p=0,
\end{align*}
for $g_n(x,p)=\text{tr}\left(M_n(x,p)^\dagger\rho M_n(x,p)\right)$.}
and rearranging yields 
\begin{align*}
   H_{\tau,\rho}(x,p)&=\cfrac{\text{tr}\left(\rho e^{-\tau((\hat{X}-x)^2+(\hat{P}-p)^2)}\right)}{\int_{\mathbb{R}^2}\text{tr}\left(\rho e^{-\tau((\hat{X}-x')^2+(\hat{P}-p')^2)}\right)\text{d}x'\text{d}p'},
\end{align*}
which is precisely \eqref{eq:1}. The numerator is simply the quantum expectation of the Gaussian semigroup $e^{-\tau((\hat{X}-x)^2+(\hat{P}-p)^2)}$. This can be translated into an expectation over phase-space against the Wigner distribution $W_\rho$ through the \textit{Weyl correspondence} \cite{anderson1969weyl} 
\begin{align*}
  &\text{tr}\left(\rho e^{-\tau((\hat{X}-x)^2+(\hat{P}-p)^2)}\right) =\\ &\int_{\mathbb{R}^2}W_\rho(x',p')e_{\star}^{-\tau((x'-x)^2+(p'-p)^2)}\text{d}x'\text{d}p',
\end{align*}
where $e_\star$ is the $\star$-exponential, defined by replacing the ordinary multiplication in the power series of the exponential with the \textit{Moyal product} \cite{ANDERSON1972423,zworski2012semiclassical}. As it turns out, this is \textit{still} proportional to a bivariate Gaussian \cite{curtright2013concise}, since 
\begin{align*}
    e_{\star}^{-\tau((x'-x)^2+(p'-p)^2)}=\frac{1}{\cosh(\hbar \tau)}e^{-\frac{1}{2\sigma_{\tau}^2}((x'-x)^2+(p'-p)^2)}.
\end{align*}
Thus, $H_{\tau,\rho}(x,p)$ is simply 
\begin{align*}
   H_{\tau,\rho}(x,p) = \cfrac{(W_\rho\ast g_{\sigma_{\tau}})(x,p)}{\int_{\mathbb{R}^2}(W_\rho\ast g_{\sigma_{\tau}})(x',p')\text{d}x'\text{d}p'},
\end{align*}
where the denominator is $1$. This is  \eqref{eq:conv}. Taking the limit as $\tau\rightarrow \infty$, we obtain the Husimi function
\begin{align*}
    H_{\infty,\tau}:=\lim_{\tau\rightarrow\infty}H_{\tau,\rho} = 
    W_\rho\ast g_{\sqrt{\hbar/2}},
\end{align*}
as claimed in the introduction. 

The above paradigm extends verbatim into \eqref{eq:Z}, by formally replacing $\hat{X}$ and $\hat{P}$ with an arbitrary collection of quantum observables $\hat{A}_1,\ldots,\hat{A}_n$ acting on a Hilbert space.
Accordingly, \eqref{eq:Z} is interpreted as the probability of recording average values $a_1,\ldots,a_n$, conditioned on observing a stream of constant outcomes from weakly measuring the observables $\hat A_1,\ldots,\hat A_n$, respectively.

As for \eqref{eq:general}, we first recall that the Weyl functional calculus \cite{anderson1969weyl}
establishes a mapping
\begin{align*}
    g \mapsto \hat G=\cfrac{1}{(2\pi)^n}\iint_{\mathbb R^n\times \mathbb R^n} g(a')e^{i\xi\cdot (\hat A-a')}\text{d}\xi\text{d}a',
\end{align*}
from functions $g$ on the generalized phase space $\mathbb R^n$, to operators $\hat G$ on the Hilbert space; $g$ is referred to as the {\em Weyl symbol} of $\hat G$ with respect to $\hat{A}_1,\ldots,\hat{A}_n$. Much like the case of $\hat X$ and $\hat P$, the Weyl correspondence satisfies
\[
\text{tr}(\rho \hat G)=\int_{\mathbb{R}^n} W_\rho(a')g(a')\text{d}a',
\]
for all quantum states $\rho$, with $W_\rho$ being the generalized Wigner distribution
\[
W_\rho(a)=\frac{1}{(2\pi)^n}\mathcal F \left(
\text{tr}(\rho e^{i\xi\cdot \hat A})
\right)(a),
\]
where $\mathcal F$ denotes the Fourier transform with respect to $\xi$ and $\xi\cdot \hat A:=\xi_1\hat A_1+\ldots+\xi_n\hat A_n$.
Provided a symbol exists for the Gaussian semigroup
$e^{-\tau{\sum_{k=1}^n(\hat{A}_k-a_k)^2}}$, the integral representation \eqref{eq:general} follows.

We conclude the Letter by showing that our paradigm also recovers the Husimi function for the spin-$1/2$ particle. To this end, consider the spin$-1/2$ operators 
\begin{align*}
    \hat{A}_1 =\hat{S}_x,~\hat{A}_2=\hat{S}_y,~\hat{A}_3=\hat{S}_z,
\end{align*}
where 
\begin{align*}
    \hat{S}_x=\begin{bmatrix}
        0&1\\
        1&0
    \end{bmatrix},~ \hat{S}_y=\begin{bmatrix}
        0&-i\\
        i&\phantom{-}0
    \end{bmatrix},~ \hat{S}_z=\begin{bmatrix}
        1&0\\
        0&-1
    \end{bmatrix}.
\end{align*}
Using \eqref{eq:Z}, the positive probability representation
\begin{align}\nonumber
    &H_{\tau,\rho}(x,y,z) =\\ &\cfrac{\text{tr}(\rho e^{-\tau((\hat{S}_x-x)^2+(\hat{S}_y-y)^2+(\hat{S}_z-z)^2)})}{\int_{\mathbb{R}^3}\text{tr}(\rho e^{-\tau((\hat{S}_x-x')^2+(\hat{S}_y-y')^2+(\hat{S}_z-z')^2)})\text{d}x'\text{d}y'\text{d}z'}\label{eq:H}
\end{align}
follows for any given quantum spin state $\rho$ \footnote{We note a slight abuse of notation in that we use $x,y,z$ for both, as values in $\mathbb R$ as well as labels indexing the Pauli matrices.}.  We next simplify the above expression and show that the limit, as $\tau\to \infty$, is the Husimi function.

To this end, recall that $\hat{S}_k\hat{S}_\ell+\hat{S}_\ell\hat{S}_k = 2\delta_{k\ell}I$, for $k,\ell\in\{x,y,z\}$, and let $\mathbf{r}=[x,y,z]$, $r=\|\mathbf r\|$. Then, \eqref{eq:H} becomes
\begin{align}
    \cfrac{e^{-\tau r^2}\text{tr}(\rho e^{2\tau(x\hat{S}_x+y\hat{S}_y+z\hat{S}_z)})}{\int_{\mathbb{R}^3}e^{-\tau r'^2}\text{tr}(\rho e^{2\tau(x'\hat{S}_x+y'\hat{S}_y+z'\hat{S}_z)})\text{d}x'\text{d}y'\text{d}z'}.\label{eq:new}
\end{align}
Defining $\mathbf{r}_\rho =\text{tr}(\rho [\hat S_x,\,\hat S_y,\,\hat S_z])$ to be the Bloch vector for the state $\rho$, one has that $\text{tr}(\rho e^{2\tau(x\hat{S}_x+y\hat{S}_y+z\hat{S}_z)})$ equals
\begin{align*}
\cosh(2\tau r)+\frac{\mathbf{r}\cdot\mathbf{r}_\rho }{r}\sinh(2\tau r).
\end{align*}
Substituting in \eqref{eq:new}, the integral with the $\sinh$ term vanishes, and
the denominator simplifies to
\begin{align*}
    \left(\frac{\pi}{\tau}\right)^{3/2}(2\tau+1)e^\tau,
\end{align*}
which is independent of the state $\rho$.
The expression in \eqref{eq:new} now becomes
\begin{align*}\nonumber
    &H_{\tau,\rho}(\mathbf{r}) =\frac{e^{-\tau(1+ r^2)}(\cosh(2\tau r)+(\frac{\mathbf{r}\cdot \mathbf{r}_\rho}{r} )\sinh(2\tau r))}{(\pi/\tau)^{3/2}(2\tau+1)}.
    \end{align*}
Rewriting the above as
    \begin{align*}
    \frac{\sqrt{\tau/\pi}}{c_\tau} ( e^{-\tau(r-1)^2} (1+\mathbf{n}\cdot \mathbf{r}_\rho)  + e^{-\tau(r+1)^2}(1-\mathbf{n}\cdot \mathbf{r}_\rho) )
\end{align*}
where $\mathbf{n}=\mathbf{r}/r$ and $c_\tau=\pi(4+2/\tau)$,
it is readily seen that the limit, as $\tau\to\infty$, vanishes outside of the shell of $r=1$, so that
\[
H_{\infty,\rho}(\mathbf{r}) = \frac{1}{4\pi} (1+\mathbf{n}\cdot \mathbf{r}_\rho) \delta(r=1).
\]
This is the Husimi function for the spin-$1/2$ particle \cite{xu2025wehrl}.


In the setting of \eqref{eq:general}, it is seen that the finite-time Husimi function $H_{\tau,\rho}$ is a smearing of the trivariate spin-$1/2$ Wigner distribution \cite{chandler1992quasi}. Specifically, we recall that the Weyl correspondence establishes a direct relation for functions of linear combinations of the operators, in that 
\begin{align*}
(a x' +by' + cz')^k \mapsto (a\hat S_x+b\hat S_y + c\hat S_z)^k.
\end{align*}
As a result,
\begin{align*}
    H_{\tau,\rho}(\mathbf{r})=\left(\frac{\tau}{\pi}\right)^{3/2}\frac{e^{-\tau(1+ r^2)}}{(2\tau+1)}\int_{\mathbb{R}^3}W_\rho(\mathbf{r}')e^{2\tau(\mathbf{r}\cdot\mathbf{r}')}\text{d}\mathbf{r}'.
\end{align*}
Thus, the natural smearing function that wipes out all negative mass from the Wigner distribution for the spin-$1/2$ particle is simply 
\begin{align*}
    \cfrac{g(\tau,\mathbf{r},\mathbf{r}')}{Z_{\tau,\rho}} = \left(\frac{\tau}{\pi}\right)^{3/2}\frac{e^{-\tau(1+ r^2)}}{(2\tau+1)}e^{2\tau(\mathbf{r}\cdot\mathbf{r}')}.
\end{align*}
We remark that similar steps can be carried out for just two Pauli matrices, in which case one still obtains a well-defined class of positive representations, albeit no longer informationally complete \cite{schwonnek2020wigner}. 


Finally, when the paradigm is applied to a single observable \footnote{This trivially extends to a collection of \textit{commuting} quantum observables.}, say $\hat{A}$, the Husimi function manifestly reduces to the distribution of projective measurement outcomes associated with $\hat{A}$, in agreement with Born's rule for quantum measurement \cite{born1926quantum}, \footnote{Note the collapse of the support of $H_{\tau,\rho}$ to the spectrum of $\hat{A}$, spec($\hat{A}$).},
\begin{align*}
\cfrac{\text{tr}(\rho e^{-\tau(\hat{A}-a)^2})}{\int \text{tr}(\rho e^{-\tau(\hat{A}-a)^2})\text{d}a} \xrightarrow{\tau\rightarrow \infty}  \int_{\text{spec}(A)} \hspace*{-5pt}\delta(a-\lambda)\text{tr}(\rho|\lambda\rangle\langle \lambda|)\text{d}\lambda.
\end{align*}  Thus, the paradigm presented herein demonstrates that the Husimi function is a \textit{natural extension of Born's rule for quantum measurement to the setting of non-commuting observables}.

When the observables don't commute, the support cannot simply collapse to the spectrum of either operator, and converges instead (when the limit exists) to more intricate shapes that can be related to both the singular supports of the generalized Wigner distributions \cite{schwonnek2020wigner,sabbagh2025particle}, as well as the topological structure of generalized coherent states \cite{perelomov1972coherent,perelomov1977generalized}. In the spin case, the support of the Husimi function, when carried out for $2$ or $3$ Pauli matrices, is seen to coincide with the singular support of the Wigner distribution; that of a circle or a sphere, respectively, as predicted in \cite{sabbagh2025particle}, providing for the first time a probabilistic interpretation of the positive regions of the spin Wigner distributions studied in \cite{cohen1986joint,sabbagh2025particle, chandler1992quasi,schwonnek2020wigner}.

\noindent
\emph{Acknowledgments:} Part of this research was conducted while the authors were visiting the Institute for Pure and Applied Mathematics (IPAM), which is supported by the National Science Foundation (Grant No. DMS-1925919).
Additionally, the research was supported by the NSF under ECCS-2347357, the AFOSR under FA9550-24-1-0278, and ARO under W911NF-22-1-0292.
OMM was supported by the European Union's Horizon 2020 research and innovation
programme under the Marie Skłodowska-Curie Grant Agreement No. 101151140.

\appendix

\bibliographystyle{unsrt}
\bibliography{Ref}

\end{document}